\documentclass[conference]{IEEEtran}

\ifCLASSINFOpdf

\else

\fi

\usepackage{graphicx}
\usepackage{color}

\usepackage{cite}
\usepackage{url}

\usepackage[cmex10]{amsmath}

\usepackage{algorithmic}

\usepackage{array}
\usepackage{algorithmic}
\usepackage[ruled,norelsize]{algorithm2e}

\usepackage{multirow}
\usepackage[english]{babel}
\usepackage{amsmath}
\usepackage{amsthm}

\usepackage[tight,footnotesize]{subfigure}

\usepackage{mdwmath}
\usepackage{mdwtab}

\title{A Hierarchical WDM-based Scalable Data Center Network Architecture}

\author{}

\begin{document}

\author{\IEEEauthorblockN{Maotong Xu$^1$, Jelena Diakonikolas$^2$, Eytan Modiano$^3$, Suresh Subramaniam$^1$} \\
	\IEEEauthorblockA{$^1$George Washington University, Washington, DC 20052, USA \\
		$^2$The University of California Berkeley, Berkeley, CA 94720, USA\\
		$^3$Massachusetts Institute of Technology, Cambridge, MA 02139, USA \\
		Email: $^1$\{htfy8927, suresh\}@gwu.edu, $^2$jelena.d@berkeley.edu, $^3$modiano@mit.edu}
}

\maketitle

%
\IEEEpeerreviewmaketitle

\begin{abstract}

Massive data centers are at the heart of the Internet. The rapid growth of Internet traffic and the abundance of rich data-driven applications have raised the need for enormous network bandwidth. 
Towards meeting this growing traffic demand, optical interconnects have gained significant attention, as they can provide high throughput, low latency, and scalability. In particular, optical Wavelength Division Multiplexing (WDM) provides the possibility to build data centers comprising of millions of servers, while providing hundreds of terabits per second bandwidth. 

In this paper, we propose a WDM-based Reconfigurable Hierarchical Optical Data Center Architecture (RHODA) that can satisfy future Internet traffic demands. To improve scalability, our DCN architecture is hierarchical, as it groups server racks into clusters. Cluster membership is reconfigurable through the use of optical switches. Each cluster enables heavy-traffic communication among the racks within. To support varying traffic patterns, the inter-cluster network topology and link capacities are also reconfigurable, which is achieved through the use of optical space switches and Wavelength Selective Switches (WSSs). Our simulation results demonstrate that in terms of average hop distance, RHODA outperforms OSA, FatTree and WaveCube by up to $81\%$, $66\%$ and $60\%$, respectively.
\end{abstract} 

\section{INTRODUCTION}
\IEEEPARstart{I}{nternet} traffic has experienced dramatic growth in the past decade. Such growth is driving an enormous need for network bandwidth and data storage, which is being enabled by massive data centers that scale along with the Internet. Conventional electrical data center networks (e.g., FatTree~\cite{al2008scalable}, Bcube~\cite{guo2009bcube}) are built using a multi-layer approach, with identical switches at the bottom level to connect with servers or racks, and expensive and high-end switches located at the upper layers to aggregate and distribute the traffic. 

However, in the coming years, data centers will consist of millions of servers, and the bisection bandwidth of a data center network (DCN) could exceed 100~Tbps, which is beyond the capability of electronic switching. For example, Microsoft owns over one million servers and its Chicago data center alone is estimated to contain over 250,000 servers \cite{website:largedatacenter2013}. Interactive applications such as web search, social networks, and stock exchange, require low network latency. For example, the acceptable latency range for stock exchange transactions is 5-100~ms \cite{website:bitlatency2012}. Optical Wavelength Division Multiplexing (WDM) is a promising technology for meeting the networking demand of data centers. It can support more than 100 wavelengths per fiber and 100~Gbps transmission rate per wavelength. Moreover, large-scale optical switches consume less energy per bit/s, making the network architecture scalable and energy-efficient.

Optical networks are commonly based on optical circuit switching, e.g., Micro-Electro-Mechanical Systems Switch (MEMS) and Arrayed Waveguide Grating Router (AWGR). 

MEMS is a power-driven reconfigurable optical switch with reconfiguration time on the order of a few milliseconds. OSA~\cite{ChenOSA} utilizes a central MEMS to connect all Top-of-Rack (ToR) switches. The architecture's scalability is limited by the number of ports of the central MEMS.  OGDCN~\cite{sankaran2016optical} builds a Clos network with multiple stages of MEMS. Multiple stages of MEMS improves scalability, but also introduces more power consumption and cost.

\begin{figure*}
	\centering
	\includegraphics[height=2.6in,width=6.0in]{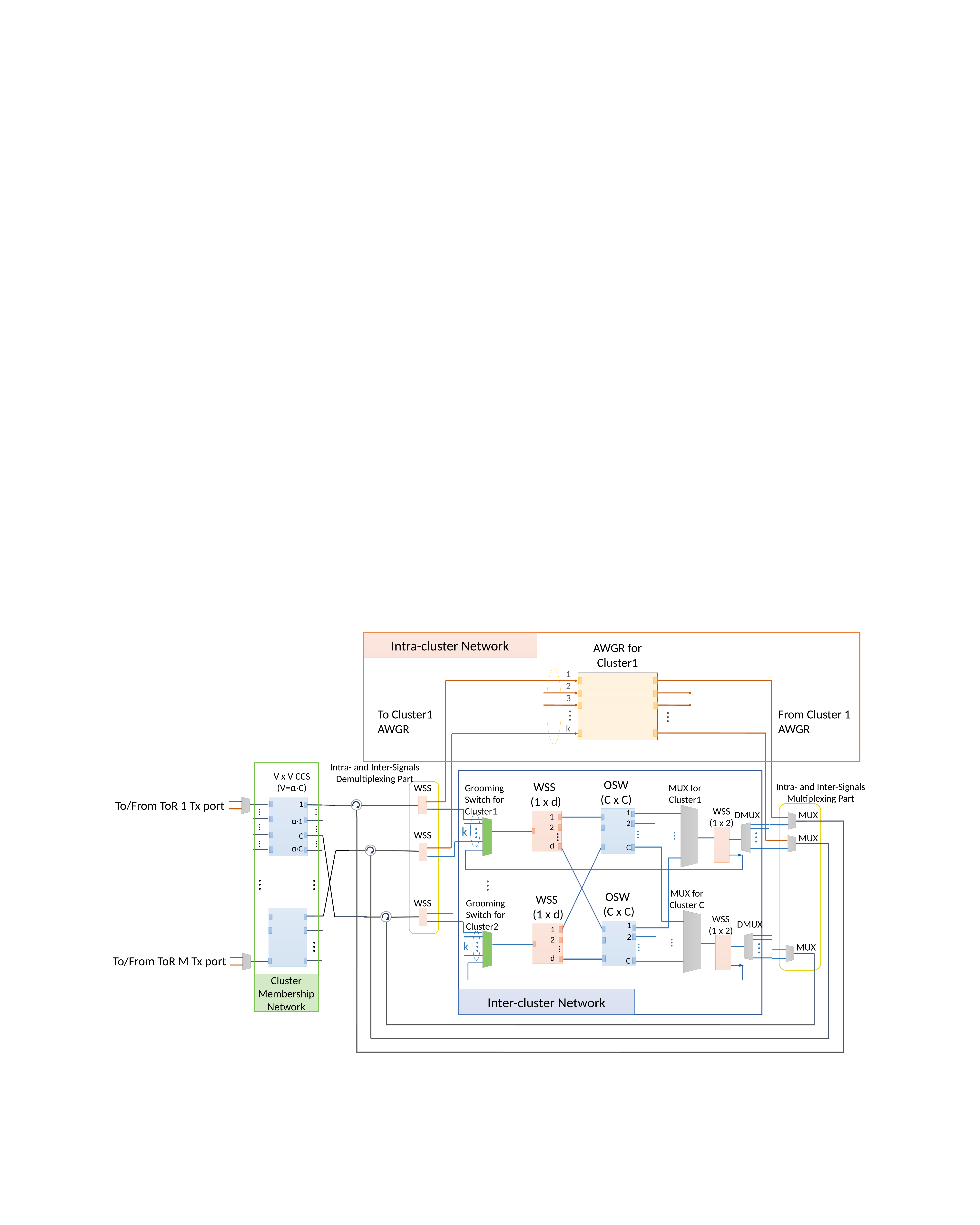}
	\caption{Two-level DCN architecture with multiple clusters of racks. \small{$M$: the number of racks; $k$: the number of racks per cluster; $d$: egress degree of a cluster; $C$: the number of clusters.}}
	\label{fig:arc}
	\vspace{-10pt}
\end{figure*}

AWGR is a passive optical device that does not require reconfiguration, and can achieve contention resolution in the wavelength domain. The cyclic routing characteristic of the AWGR allows different inputs to reach the same output simultaneously by using different wavelengths. DOS~\cite{ye2010scalable} utilizes a central AWGR to connect all ToR switches. The architecture's scalability is limited by the number of ports of the central AWGR. Petabit~\cite{xia2010petabit} improves scalability by using multiple stages of AWGR. However, a large number of tunable wavelength converters (TWC) are introduced to tune to appropriate wavelengths. PODCA \cite{xu2016podca} is highly scalable and consists entirely of passive optical devices. However, as other AWGR-based architectures, PODCA lacks reconfigurability, and it cannot configure its topology based on features of traffic patterns.

WaveCube~\cite{chen2015wavecube} achieves scalability and high-performance simultaneously by employing WSSs to provide dynamic link bandwidth. However, WaveCube is not designed for accommodating different traffic patterns. Opsquare \cite{yan2017opsquare} and HiFOST~\cite{yan2018hifost} use fast optical switches (FOS), whose reconfiguration time is a few nanoseconds, for intra-cluster communications. Opsquare also uses FOS for inter-cluster communications, while HiFOST uses Ethernet for inter-cluster communications. FOS is a protoype implementation available only in laboratory settings, and is not ready to use in current data centers. 

Our solution is to utilize a combination of both MEMS and AWGR, which provide us with both reconfigurability and low power consumption. In this paper, we introduce a new WDM-based Reconfigurable Hierarchical Optical Data Center Architecture (RHODA). The architecture accommodates massive numbers of servers and varying traffic patterns. It is hierarchical and achieves high capacity through reconfigurable clustering of racks of servers, wherein separate resources are dedicated to communication within and between the clusters. The clustering of racks and reconfigurability of cluster membership and inter-cluster topology make the new architecture  suitable for different traffic patterns. Our simulation results demonstrate that, in terms of average hop distance, RHODA outperforms OSA, FatTree and WaveCube by up to $81\%$, $66\%$ and $60\%$, respectively.

The rest of this paper is organized as follows: Section~\ref{sec:designdetails} describes the details of our architecture. Section~\ref{sec:algorithm} presents algorithms for cluster configuration, wavelength assignment, and routing.  Section~\ref{sec:evaluation} describes the performance evaluation of the architecture, in comparison with other architectures. Section~\ref{sec:PC} presents comparison results in terms of power and cost. Section~\ref{sec:conclusion} concludes the paper.

\vspace{-5pt}
\section{RHODA Design Details}\label{sec:designdetails}

\begin{figure*}
	\centering
	\includegraphics[height=1.6in,width=5.6in]{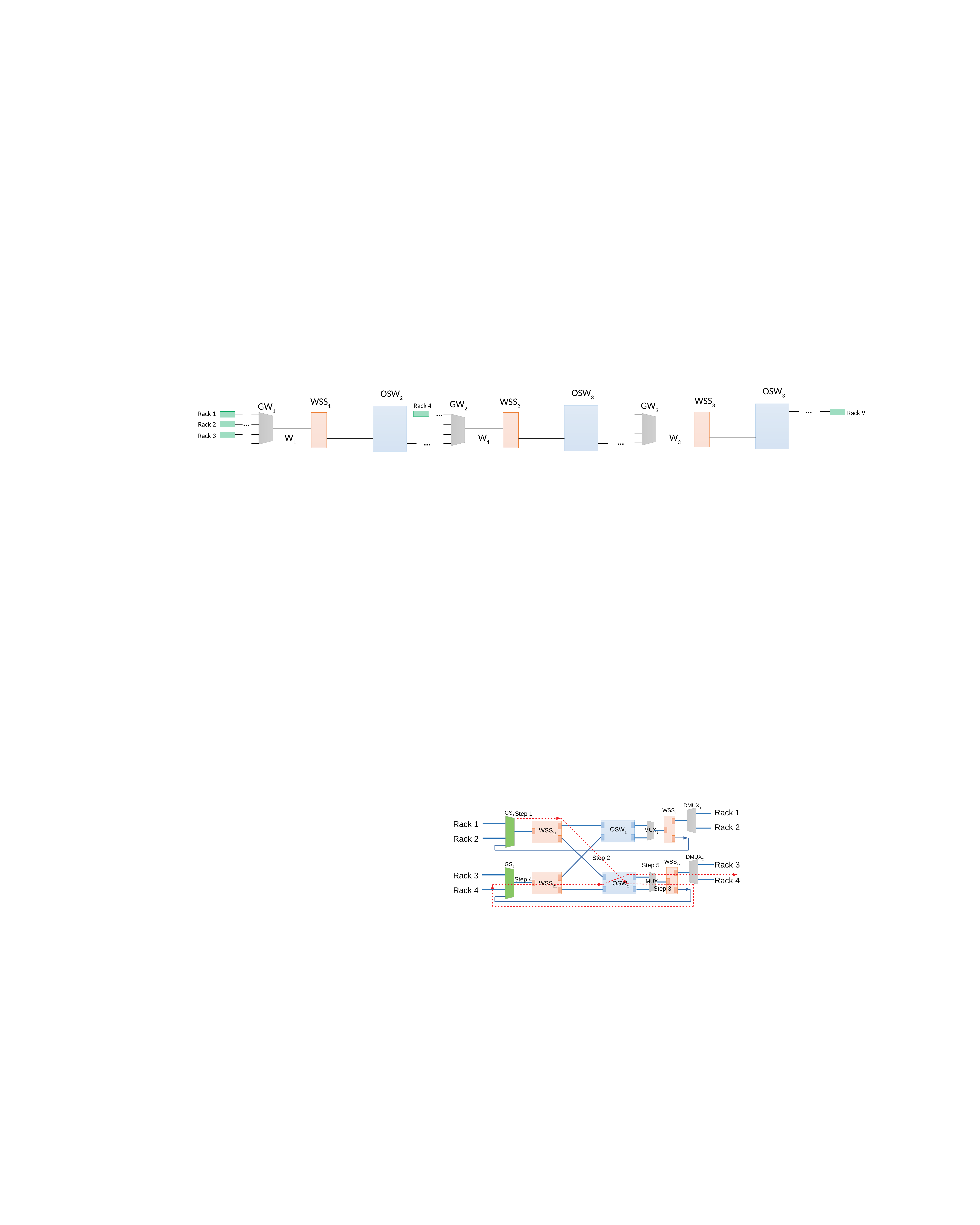}
	\caption{Inter-cluster routing example.}
	\label{fig:example}
	\vspace{-12pt}
\end{figure*}

The DCN consists of $N$ servers grouped into $M$ racks, so that there are $m=N/M$ servers per rack. Each rack contains a ToR switch used for both electronic switching of the packets within a rack and for communication with other racks. Each ToR is directly connected to all of the $m$ servers within the rack. For communication to and from other ToRs, each ToR contains $T^{\mathrm{intra}} + T^{\mathrm{inter}}$ tunable transceivers (TRXs). 
Fig.~\ref{fig:arc} depicts the proposed two-level hierarchical DCN architecture, in which racks are partitioned into $C$ clusters, with each cluster comprised of $k = M/C$ racks. 

As Fig.~\ref{fig:arc} illustrates, RHODA consists of five parts, i.e., cluster membership network (CMN), intra- and inter- signals demultiplexing part, intra-cluster network, inter-cluster network, and intra- and inter- signals multiplexing part. 

To support varying data traffic patterns~\cite{ChenOSA}, we let cluster membership be reconfigurable. This is achieved by the CMN using a set of $k/\alpha$ $V{\times}V$ cluster configuration switches (CCS), where $V=\alpha{\cdot}C$ and $\alpha$ is a positive integer parameter that is chosen as a trade-off between the cost and complexity of the CMN and the reconfiguration flexibility. First, all the wavelengths from a ToR are multiplexed onto a single fiber before being connected to the CMN. The $p$th output port of a CCS is connected to cluster $\lceil\frac{p}{\alpha}\rceil$. At most $\alpha$ racks that are connected to a CCS can be configured to the same cluster. When $\alpha = k$, the CMN is fully flexible and any rack can be configured to any cluster and the CMN is a single large $M\times M$ switch, and when $\alpha = 1$, the CMN is least flexible as only one out $C$ racks connected to the same CCS can be configured to the same cluster, but the CMN consists of $k$ small ($C \times C$) switches.

Next, in the signals demultiplexing part, the signals at an output port of a CCS are separated onto two fibers, with one fiber carrying the wavelengths used for intra-cluster communication, and the other fiber carrying the wavelengths used for inter-cluster communication.

Since most of the heavy communication in a data center is carried over small subsets of ToRs~\cite{Roy15} (and these ToRs\footnote{We use the terms ``rack" and ``ToR" interchangeably in the paper.} would ideally be configured to the same cluster), in intra-cluster network, RHODA equips each cluster with a $k{\times}k$ AWGR to support large amounts of intra-cluster traffic. The AWGR is a passive optical device that routes wavelengths from input ports to output ports in a cyclic manner. In particular, wavelength $\lambda_i$ appearing on port $p$ is routed to port $q=[(i+p-2)  \mod \ k]+1, \ 1{\leq}i{\leq}W^{\mathrm{intra}}$, where $W^{\mathrm{intra}}$ is the number of wavelengths available for intra-cluster communication. 

In the inter-cluster network, each cluster can be considered as the smallest communication element. Flows from racks are merged (using optical-to-electrical-to-optical conversion) by grooming switches (GSs) to reduce the number of wavelengths needed. The communication graph between clusters (i.e., the cluster logical topology) is then determined by $C$ $1{\times}d$ wavelength selective switches (WSSs) and $d$ $C{\times}C$ optical switches (OSWs). In particular, each cluster can send signals to up to $d$ other clusters. Demultiplexers (DMUXs) split signals carried by different wavelengths. A signal carried on wavelength w is forwarded to the $\lceil\frac{w}{k}\rceil$th port of the DMUX. 

In the signals multiplexing part, the signals from intra- and inter-cluster networks are multiplexed onto one fiber and multiplexed signals are routed to the corresponding rack. Also, circulators are introduced to enable bidirectional communication over a fiber, which allows more efficient use of CCS ports. Circulator is a three-port device that has a switch port, send port, and receive port, and can enable directional transmissions.

In the following, we use a small example to illustrate flow routing in the inter-cluster network. As shown in Fig.~\ref{fig:example}, the data center has four racks, and rack $i$ is configured to cluster $\lceil\frac{i}{2}\rceil$. Suppose rack1 and rack2 send two flows to rack4, and each flow needs 1 Gbps bandwidth. Suppose there are two wavelengths available and each wavelength has 100 Gbps bandwidth. Five steps are needed to route flows from rack1 and rack2 to rack4. 

\begin{enumerate}
	\item First, rack1 and rack2 send flows to $GS_1$. $GS_1$ receives data from the two flows, and sends a new flow carried on a single wavelength (e.g., $W_1$).
	\item The flow is forwarded to $OSW_2$ via the second port of $WSS_{11}$.
	\item Since $DMUX$ is a passive optical device, $DMUX_2$ can only forward a signal carried on $W_2$ to rack4. Therefore, the flow's wavelength needs to be converted from $W_1$ to $W_2$, so the flow is routed to $GS_2$ to change the wavelength.
	\item $GS_2$ receives the flow and sends a new flow carried on $W_2$.
	\item Finally, the flow is routed to rack4 via $OSW_2$, $MUX_2$, $WSS_{22}$, and $DMUX_2$.
\end{enumerate}
With a 128-port AWGR and a 320-port MEMS switch, RHODA can accommodate more than 2.6 million servers. We investigate the scalability of the RHODA architecture in Section~\ref{sec:scalability}.


\vspace{-5pt}
\section{Cluster Configuration, Wavelength Assignment, and Routing Algorithms}\label{sec:algorithm}

Given traffic demands between racks~\footnote{Estimating traffic demands is outside the scope of this paper, but can be done using the technique in~\cite{Farrington10}, for instance.}, we present simple algorithms for configuring cluster membership, routing intra-cluster flows, and configuring inter-cluster topology, routing inter-cluster flows, and computing the wavelength assignment. We also discuss wavelength assignment schemes together with intra- and inter-cluster routing algorithms. The reconfiguration of a DCN with millions of servers must be completed within a few milliseconds. Thus, instead of seeking optimal/approximation algorithms, we propose simple, but effective heuristic algorithms, but we believe that there is room for further research on this issue.

\vspace{-5pt}
\subsection{Cluster Membership Configuration} 
Grouping racks with massive mutual traffic into a cluster can reduce traffic congestion and high latency that results from inter-cluster multi-hop communication. We first sort mutual traffic between each pair of racks in non-increasing order, and place the rack pair with the $l^{\mathrm{th}}$ largest mutual traffic into cluster $(l \mod \ C)+1$, where $C$ is the total number of clusters.  

\vspace{-5pt}
\subsection{Intra-cluster Topology Configuration and Routing}  
As an AWGR can simultaneously route traffic between all pairs of input-output ports, the routing constraints come from the finite number of transceivers on each rack. Each rack can send or receive from $T^{intra}$ racks with one hop. 

For all racks to be reachable from each other, the intra-cluster logical topology needs to be connected. Moreover, racks pairs with high mutual traffic should be separated by a small number of hops to avoid high traffic congestion due to multi-hopping.

Towards achieving these goals, we first create a cycle of racks to guarantee that the rack topology is connected; racks are connected in decreasing order of mutual traffic. Then, we sort all rack pairs based on directed traffic, and connect them without exceeding the egress degree, $T^{intra}$, of any rack.

Between racks, we use the classical Dijkstra's algorithm to find the shortest paths, where flows are routed in a multi-hop manner.

\vspace{-5pt}
\subsection{Inter-cluster Topology Configuration and Routing} 

The ingress/egress degree of a cluster is $d$. To make the inter-cluster topology connected, we first create a cycle of clusters. Then, we iteratively use the Hungarian algorithm~\cite{kuhn1955hungarian} to a find perfect bipartite matching to maximize total traffic of the connected edges $d-1$ times (as the initial cycle already accounted for one ingress/egress degree of the clusters). We use Dijkstra's algorithm to find the shortest paths between clusters, where flows are routed in a multi-hop manner.


\vspace{-5pt}
\section{Evaluation}\label{sec:evaluation}

As discussed in Section~\ref{sec:designdetails}, RHODA is highly reconfigurable, which means that it can reconfigure cluster membership, and intra- and inter-cluster topology to accommodate different traffic patterns. In the first two subsections, we show the benefits of introducing reconfigurability in cluster membership, and inter-cluster topology. Further, we evaluate RHODA using data traces from Facebook. In the following evaluations, we assume the following numbers for the DCN: each cluster has 16 racks ($k = 16$), the number of wavelengths on a fiber is 128, the bandwidth of a wavelength is 100 Gbps, and both $T^{intra}$ and $T^{inter}$ are equal to 2 (i.e., a total of 4 transceivers per rack). Instead of performing packet-level simulations, we conduct flow-level simulations and choose the average number of hops as our performance metric. The average number of hops can be taken as a measure of the packet latency. We also show the average load per switch (in bps) on the right-hand axis. The average switch load is given by the product of the average number of hops and the total ingress traffic divided by the number of switches, and is therefore just a multiple of the average number of hops. We also present the maximum load over switches.

\begin{figure}[!htbp]
	\centering
	
	\subfigure[]{%
		\includegraphics[height=1.3in,width=0.25\textwidth]{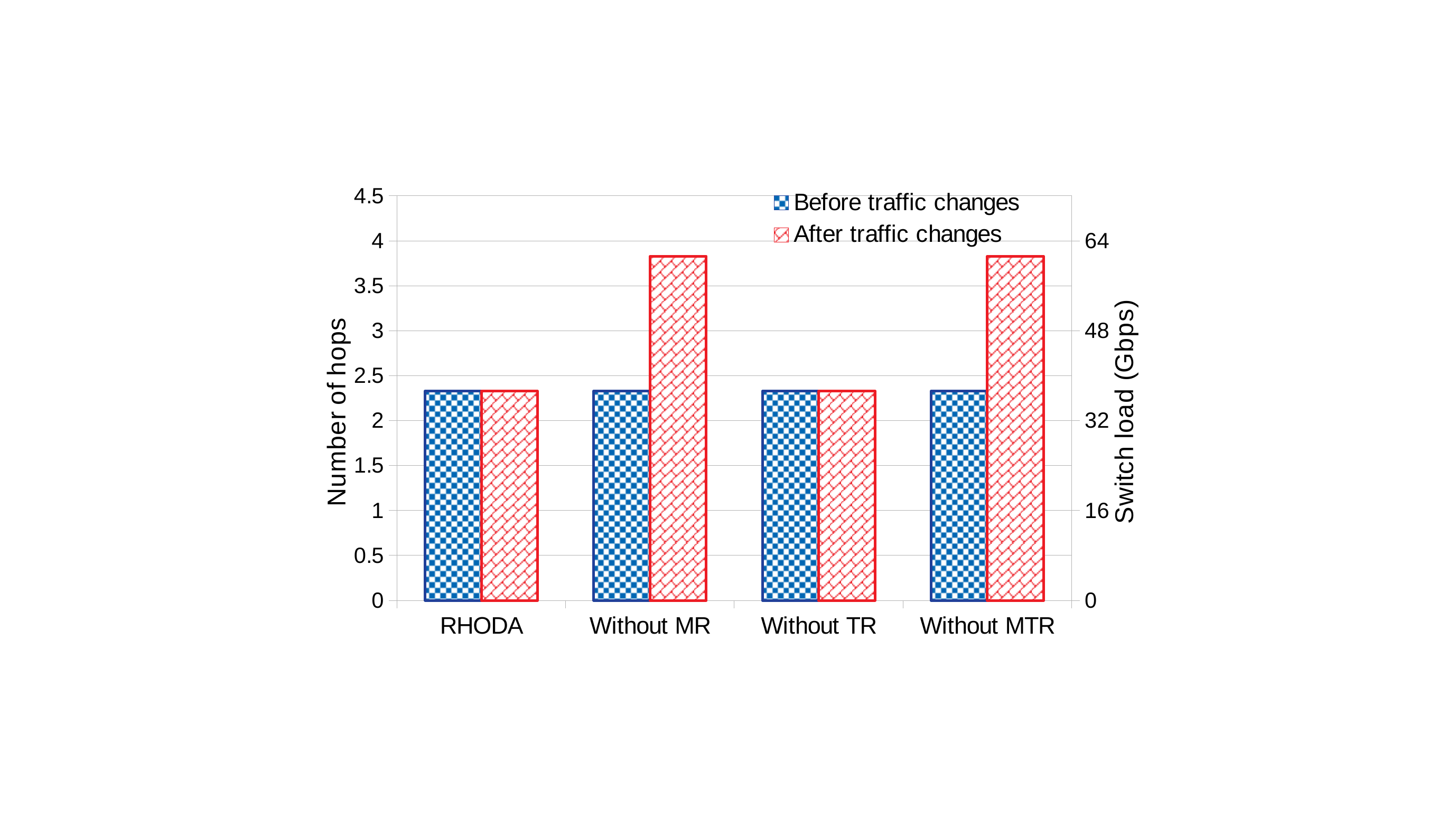}%
		\label{fig_h1}%
	}%
	~
	\subfigure[]{%
		\includegraphics[height=1.3in,width=0.25\textwidth]{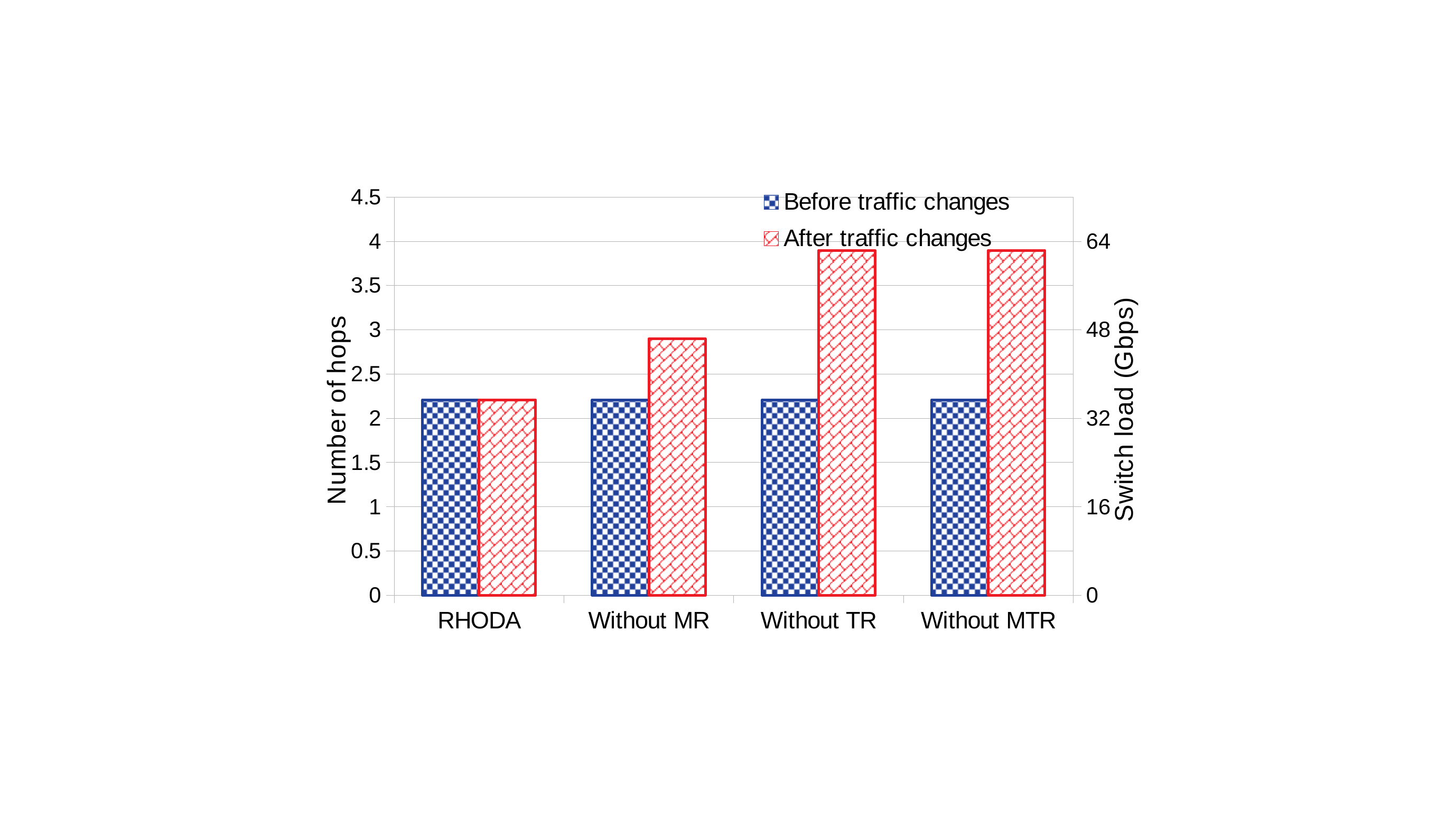}%
		\label{fig_h2}%
	}%
	~
	\vspace*{-.2cm}
	\caption{The average number of hops and switch loads with different reconfigurabilities.}
	\vspace{-10pt}
\end{figure}

\vspace{-5pt}
\subsection{Cluster Membership Reconfigurability}\label{subsec:intra}

We first show how cluster membership reconfiguration can help improve the performance of RHODA. The objective of reconfiguration is to group racks with large mutual traffic into clusters. We first evaluate RHODA on a traffic pattern with large mutual-rack traffic in a data center with 1024 racks. Such a traffic pattern is common in data centers, e.g., shuffle phase in Hadoop and Spark. The traffic pattern is generated as follows. In each iteration, randomly choose a set of $16$ racks that have not been chosen before. Then assign $\frac{16}{16}$ Gbps flow from each rack to each of the other $16$ racks in the set. The remaining $\frac{16}{16}$ Gbps of traffic generated at a rack is equally split and sent to racks outside this set. This procedure is repeated until all racks have been chosen. In this traffic pattern, there is large mutual traffic within each of the 16-rack sets but little traffic between sets. It is clear that RHODA can provide good performance if the clusters can be configured to be the sets of 16 racks with large mutual traffic.

The evaluation consists of two phases, i.e., before and after the traffic changes. In phase 1, random traffic is generated as above and RHODA is fully configured, i.e., the CCSs, OSWs, and WSSs are all configured for the traffic, and the performance is recorded. In phase 2, we regenerate another random traffic using the procedure above. Since the sets of 16 racks are randomly chosen, RHODA is now not ``matched'' to the traffic pattern. We consider four different cases to investigate which reconfigurability has the most impact for this type of traffic: RHODA with full reconfigurability, RHODA without cluster membership reconfigurability (Without MR, but with TR), RHODA without inter-cluster topology reconfigurability (Without TR, but with MR), and RHODA without cluster membership or inter-cluster reconfigurability (Without MTR).

As shown in Figure~\ref{fig_h1}, RHODA with full reconfigurability can accommodate traffic without hurting performance (the number of hops). RHODA without MR cannot accommodate this traffic pattern well as seen from the large increase in number of hops. The major reason is that without grouping racks with large mutual traffic into clusters, a large amount of traffic is routed over the inter-cluster network with increasing number of hops. However, the performance of RHODA without TR (but with MR) is comparable to that of RHODA with full reconfigurability. This is because the amount of inter-cluster traffic in this traffic pattern is small when the clusters are reconfigurable. Even though the inter-cluster topology is not well-suited to the traffic, it does not hurt the performance much. Also, we observe that the performance of RHODA without MTR is comparable to that of RHODA without MR; this further demonstrates that cluster membership reconfigurability dominates the performance for this traffic pattern. Further, after the traffic changes, the maximum switch load of RHODA without MR (1068 Gbps) is almost 4 times the maximum switch load of RHODA (271 Gbps).

\vspace{-5.5pt}
\subsection{Inter-cluster Topology Reconfigurability}\label{subsec:inter}

Inter-cluster topology and logical link bandwidth reconfigurability are achieved through OSWs and WSSs. The objective is to connect clusters with large mutual traffic using a small number of hops. Here, we randomly choose a set of $48$ racks in each iteration, and exclude these $48$ racks in subsequent iterations. At each iteration, we assign $\frac{16}{48}$ Gbps flow from a rack to each of the other $47$ racks in the set. Also, a rack has a total of $\frac{16}{48}$ Gbps traffic to other racks outside of the set of $48$ racks (split equally). In this traffic pattern, if RHODA is fully configured to support the traffic, not only is there heavy traffic within each cluster, but there is also heavy traffic within a set of 3 clusters (recall that there are 16 racks in a cluster; $48 = 3 \cdot 16$).

As shown in Figure~\ref{fig_h2}, RHODA with full reconfigurability can accommodate the traffic well. As we have discussed above, this traffic pattern has larger inter-cluster traffic than the one in Section~\ref{subsec:intra}, so inter-cluster topology reconfigurability is more important. As can be seen, RHODA without MR can provide good performance, but due to the low utilization of the intra-cluster network, more traffic has to be routed over the inter-cluster network, and this increases the average number of hops. Further, after the traffic changes, the maximum switch load of RHODA without MR (600 Gbps) is more than two times the maximum switch load of RHODA (258 Gbps), and the maximum switch load of RHODA without TR (1096 Gbps) is more than 4 times the maximum switch load of RHODA (258 Gbps).

\begin{figure}[t!]
	\centering
	\includegraphics[height=1.5in,width=2.8in]{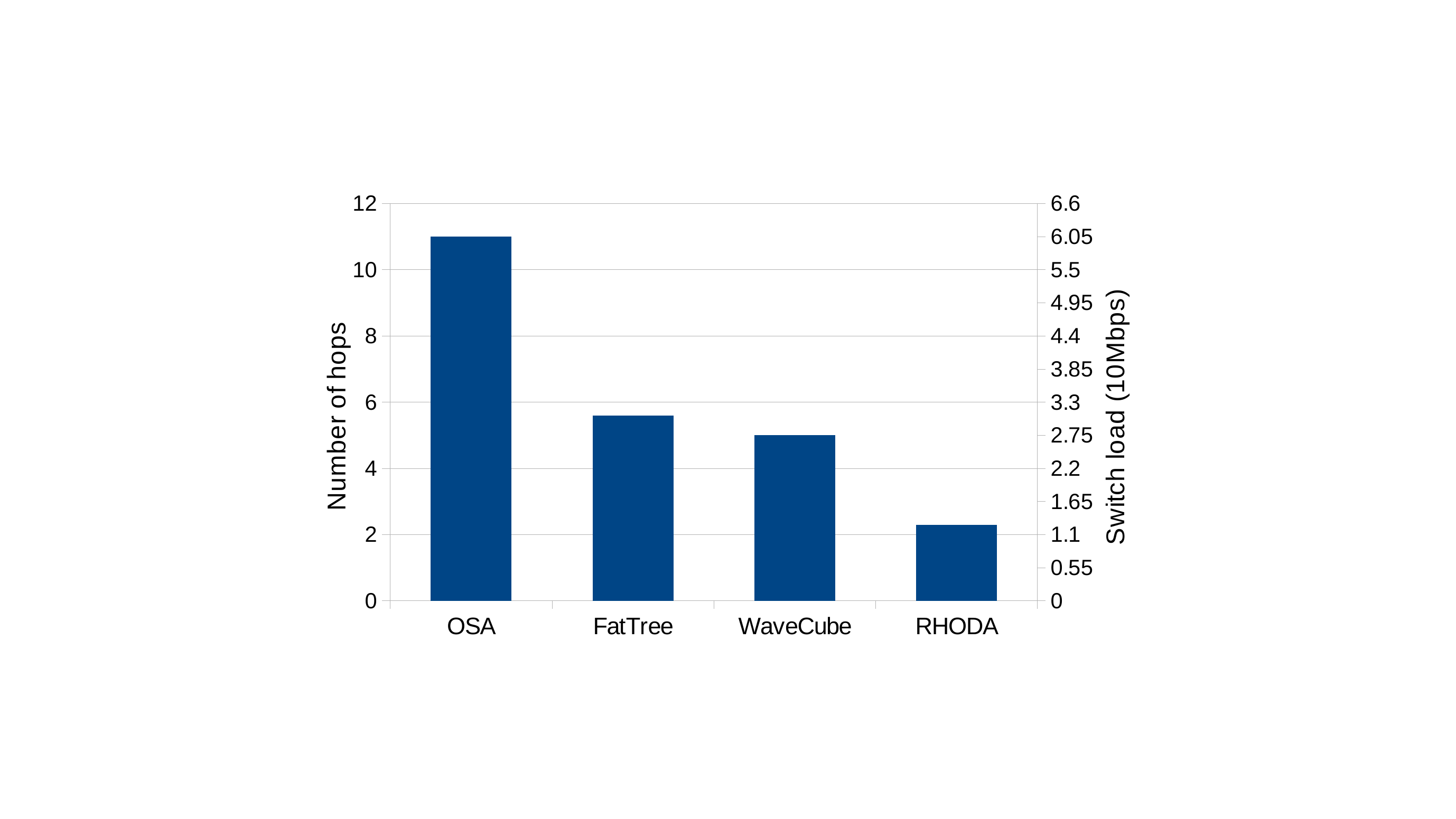}
	\caption{Performance comparisons with Facebook trace.}
	\label{fig_facebook}
	\vspace{-10pt}
\end{figure}

\vspace{-8pt}
\subsection{Evaluation on a Facebook Trace}

Traces from Facebook~\cite{Roy15} present different characteristics compared with traffic patterns from~\cite{Benson2010,ChenOSA}. Traces show that traffic is neither rack-local nor all-to-all, and heavy flow rarely happens in datacenters, e.g., more than $90\%$ flows are $900$ KBps. Also, traffic is high fan-out, i.e., each source has a large number of destinations. We extract source and destination pairs from the traces, and generate flow sizes based on the CDF of flow rates from~\cite{Roy15}.

We compare RHODA with an electrical architecture (FatTree), a configurable optical architecture (OSA), and a non-configurable optical architecture (WaveCube) in terms of the number of hops, by using traffic traces from Facebook~\cite{Roy15}. We also compare RHODA with OSA and WaveCube in terms of the number of wavelengths needed.

Figure~\ref{fig_facebook} shows the comparison results. 
As FatTree is a three-level hierarchical architecture, the maximum hop distance is $6$. When the network is large, flows in FatTree can achieve smaller hop distances, compared with OSA. In FatTree, a switch sends flows generated from multiple racks, e.g., an edge switch sends flows generated from ${\beta}/2$ racks to aggregation switches, where $\beta$ is the number of ports on a switch. FatTree lacks topology reconfigurability, which means it cannot directly connect racks with large traffic. The result shows that RHODA has the best performance under Facebook traces. In terms of average hop length, RHODA outperforms OSA, FatTree and WaveCube by up to $81\%$, $66\%$ and $60\%$, respectively.

\vspace{-5pt}
\section{Power and Cost Comparison}\label{sec:PC}

\begin{table}
	\centering
	\caption{Power consumption and cost of components in different DCN architectures}
	\label{power_cost}
	\scalebox{0.7}{
	\begin{tabular}{|c|c|c|c|c|c|}
		\hline
		\multirow{3}{*}{\textbf{Component}} & \multirow{3}{*}{\textbf{\begin{tabular}[c]{@{}c@{}}Power\\ (W)\end{tabular}}} & \multirow{3}{*}{\textbf{\begin{tabular}[c]{@{}c@{}}Cost\\ (\$)\end{tabular}}} & \multirow{3}{*}{\textbf{FatTree}} & \multirow{3}{*}{\textbf{WaveCube}} & \multirow{3}{*}{\textbf{RHODA}} \\ 
		&  &  &  &  & \\
		&  &  &  &  & \\ \hline \hline
		
		\textbf{Electronic} & \multicolumn{5}{c|}{} \\ \hline
		
		\begin{tabular}[c]{@{}c@{}}Switch per \\ port \cite{chen2015optical} \end{tabular} & 50 & 3465 & $N_{FT}$ & 0 & 0 \\ \hline
		\begin{tabular}[c]{@{}c@{}} NIC \cite{chen2015optical} \end{tabular} & 4 & 1125 & $N_{FT}$ & 0 & 0 \\ \hline \hline
		
		\textbf{Optical} & \multicolumn{5}{c|}{} \\ \hline
		
		\begin{tabular}[c]{@{}c@{}}SFP transceiver \\ \cite{fiorani2014energy} \\ \end{tabular} & 1 & 45 & 0 & $N_{WC}$ & $N_R + M + N_{GS}$ \\ \hline
		
		\begin{tabular}[c]{@{}c@{}}Circulator \cite{chen2015optical} \\ \end{tabular} & 0 & 105 & 0 & $\lceil\log_2{M}\rceil{\cdot}M$ & $M$ \\ \hline
		
		\begin{tabular}[c]{@{}c@{}}MEMS per \\ port \cite{singla2010proteus} \end{tabular} & 0.24 & 500 & 0 & 0 & $M+d{\cdot}C$ \\ \hline
		
		\begin{tabular}[c]{@{}c@{}} AWGR per \\ port \cite{chen2015optical} \end{tabular} & 0.0 & 15 & 0 & $0$ & $M$ \\ \hline
		
		Coupler \cite{chen2015optical} & 0.0 & 195 & 0 & $M$ & $0$ \\ \hline
		
		\begin{tabular}[c]{@{}c@{}} (DE)MUX \\ per port \cite{website:DEMUX} \end{tabular} & 0.0 & 40 & 0 & $2{\cdot}{N_{WC}}$ & $(2{\cdot}k + d + 1){\cdot}C$ \\ \hline
		
		\begin{tabular}[c]{@{}c@{}}WSS per \\ port \cite{chen2015optical} \end{tabular} & 15 & 1245 & 0 & $\lceil\log_2{M}\rceil{\cdot}M$ & $(d + 2){\cdot}C + 2{\cdot}M$ \\ \hline
	\end{tabular}}
	\vspace{-10pt}
\end{table}

In this section, we compare RHODA with an electrical (Fat-tree) and an optical (WaveCube) DCN architecture in terms of power consumption and capital expenditure (CapEx). To get the number of transceivers on each switch, we assume that the datacenter serves uniform traffic. More specifically, there are $M$ racks in the datacenter, and each rack sends $R$ Gbps traffic, and a total fraction $\alpha$ of this traffic is destined to the other $M-1$ racks (with the remaining traffic staying within the rack). Table~\ref{power_cost} summarizes the power consumption and costs of various components (along with the source of the data). All traffic is assumed to be in units of Gbps below.


For the electrical DCNs, the total power consumption of the DCN is the sum of the power consumed by all the switches. Some papers have already given detailed analysis on these architectures and deduced the the number of components for a particular size of the DCN~\cite{al2008scalable, yao2014comparative}.

For Fat-tree topology~\cite{al2008scalable}, we know that all switching elements are identical, enabling us to leverage cheap commodity parts for all of the switches in the communication. Let us assume that $\beta$ is the number of ports per commodity switch in the network. The number of edge switches, aggregation switches, and core switches is ${\beta}^2/2$, ${\beta}^2/2$, and ${\beta}^2/4$, respectively. To get the power consumption and cost of transceivers, we first calculate switch load, and then get the traffic sent from each port. Then, we assume that power consumption and cost, respectively of a transceiver needed equals power consumption cost, respectively, of a $100$ Gbps transceiver multiplied by the ratio of the line rate needed between the transceivers and $100$ Gbps. In FatTree, each switch has upward (towards core switches) and downward (to hosts) traffic. In the following, we present equations to calculate traffic on each switch.

Total traffic from an edge switch to aggregation switches equals
\begin{equation*}
	T^{up}_{edge} = \frac{\alpha{\cdot}R{\cdot}M-\frac{\alpha{\cdot}R}{M-1}{\cdot}({\beta}/2-1){\cdot}M}{{\beta}^2/2},
\end{equation*}
where $\alpha{\cdot}R{\cdot}M$ is total amount of traffic from all racks, and $\frac{\alpha{\cdot}R}{M-1}{\cdot}({\beta}/2-1){\cdot}M$ is the amount of traffic without passing through aggregation switches.

Total traffic from an aggregation switch to core switches is 
\begin{equation*}
	T^{up}_{agg} = \frac{\alpha{\cdot}R{\cdot}M-\frac{\alpha{\cdot}R}{M-1}{\cdot}({\beta}^2/2-1){\cdot}M}{{\beta}^2/2},
\end{equation*}
where $\frac{\alpha{\cdot}R}{M-1}{\cdot}({\beta}^2/2-1){\cdot}M$ is the amount of traffic without passing through core switches.

Total traffic transmitted by a core switch is 
\begin{equation*}
	T_{core} = \frac{\alpha{\cdot}R{\cdot}M-\frac{\alpha{\cdot}R}{M-1}{\cdot}({\beta}^2/2-1){\cdot}M}{{\beta}^2/4}.
\end{equation*}

Total traffic from an aggregation switch to edge switches is 
\begin{equation*}
	T^{down}_{agg} = \frac{\alpha{\cdot}R{\cdot}M-\frac{\alpha{\cdot}R}{M-1}{\cdot}({\beta}/2-1){\cdot}M}{{\beta}^2/2}
\end{equation*}

Total traffic from an edge switch to hosts is
\begin{equation*}
	T^{down}_{edge} = \alpha{\cdot}R{\cdot}\frac{\beta}{2}.
\end{equation*}

So, the total number of $100$ Gbps transceivers needed is 
\begin{equation*}
	N_{FT} =  {\frac{{\beta}^2}{2}}\left(\frac{T^{up}_{edge}+T^{up}_{agg}+T_{core}/2+T^{down}_{agg}+T^{down}_{edge}}{100}\right).
\end{equation*}

Our study changes the topology size from thousands to millions of racks. Even though the number of ports at each commodity edge switch is limited, we calculate the power and cost of switches by multiplying the unit price per port with the number of ports~\cite{ChenOSA}. The unit price of ToR switch per port is \textdollar $3465$, and the power of the ToR per port can be calculated as $50$ Watts. If we need $24$ ports of ToR switch to support around $10,000$ servers, the price of a matching ToR switch is $3465{\times}24$ = \textdollar $83160$, and power is $1200$ Watts.

In the optical DCNs, $M$ is the total number of racks in the DCN, and each rack is assumed to have 64 servers, which have a maximum transmission rate of 10 Gbps each. Racks in WaveCube are connected based on the HyperCube topology. WaveCube consists of WSSs, couplers, multiplexer/demultiplexers, and circulators. The degree of WaveCube is $\log_2M$. The average hop length between two racks of an $n$-cube is given by~\cite{harary1988survey}: 
\begin{equation*}
	\bar{H}_{WaveCube} = \frac{n{\cdot}2^{n-1}}{2^n-1}.
\end{equation*}
So, the average switch load equals 
\begin{equation*}
	T_{WaveCube} = \alpha{\cdot}R{\cdot}\bar{H}_{WaveCube}.
\end{equation*}
The number of transceivers needed is $N_{WC}=M{\cdot}\max(T_{WaveCube}/100, \log_2M)$, where $\log_2M$ is the degree of a ToR switch, which is also the minimum number of transceivers needed.

RHODA employs a cluster membership network based on $V{\times}V$ CCSs, optical space switches based on $C{\times}C$ MEMSs, and WSSs to achieve reconfigurability. We assume that RHODA racks are connected using a unidirectional Shufflenet~\cite{hluchyj1991shuffle} topology for both the intra- and inter-cluster logical topologies. The average hop length between two racks/clusters is
\begin{equation*}
\bar{H} = \frac{b{\cdot}a^b(a-1)(3b-1)-2b(a^b-1)}{2(a-1)(b{\cdot}a^b-1)},
\end{equation*}
where $a$ is the number of egress degrees of each rack/cluster and $b$ is the number of columns of Shufflenet network~\cite{hluchyj1991shuffle}.

Suppose the average hop length between two intra-cluster racks is denoted as $\bar{H}_{intra}$. Then,
\begin{equation*}
	T^{intra}_{RHODA} = \frac{\alpha{\cdot}R}{M-1}{\cdot}(k-1){\cdot}\bar{H}_{intra}.
\end{equation*}

For inter-cluster network, we assume the cluster connects to four clusters (i.e., $d = 4$, as is typical for commercial WSSs) and the average hop length between two racks is denoted as $\bar{H}_{inter}$. The average load of a ToR switch is 
\begin{equation*}
	T^{inter}_{RHODA} = \frac{\alpha{\cdot}R}{M-1}{\cdot}(M-k).
\end{equation*}


So, the number of transceivers needed on all ToR switches is 
\begin{equation*}
N_{R} = M{\cdot}(T^{intra}_{RHODA}/100 + T^{inter}_{RHODA}/100),
\end{equation*}
where $100$ is the capacity of a wavelength in Gbps.

\begin{figure}[!htbp]
	\centering
	
	\subfigure[]{%
		\includegraphics[height=1.6in,width=0.25\textwidth]{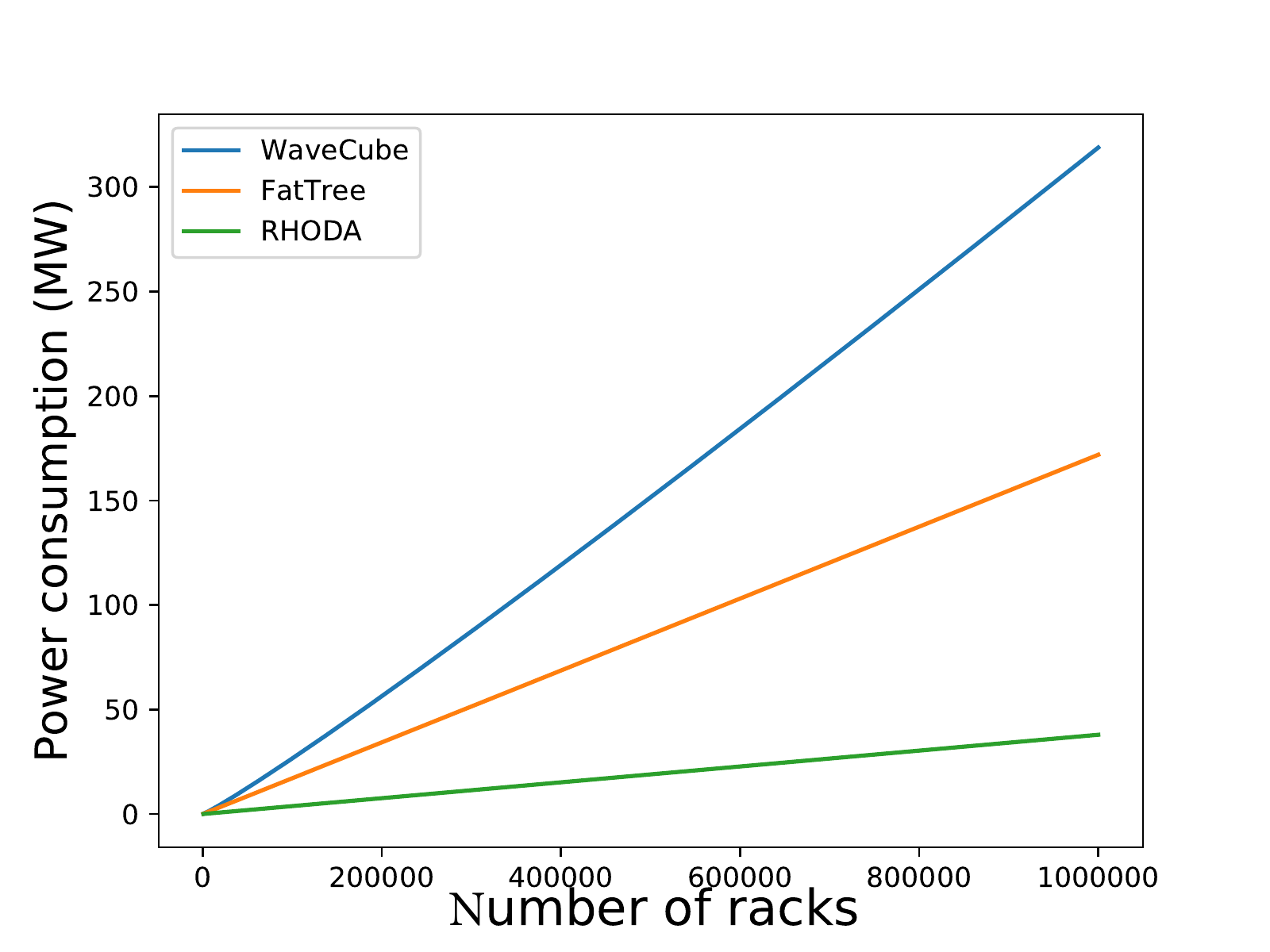}%
		\label{fig_p}%
	}%
	~
	\hspace{-10pt}
	\subfigure[]{%
		\includegraphics[height=1.6in,width=0.25\textwidth]{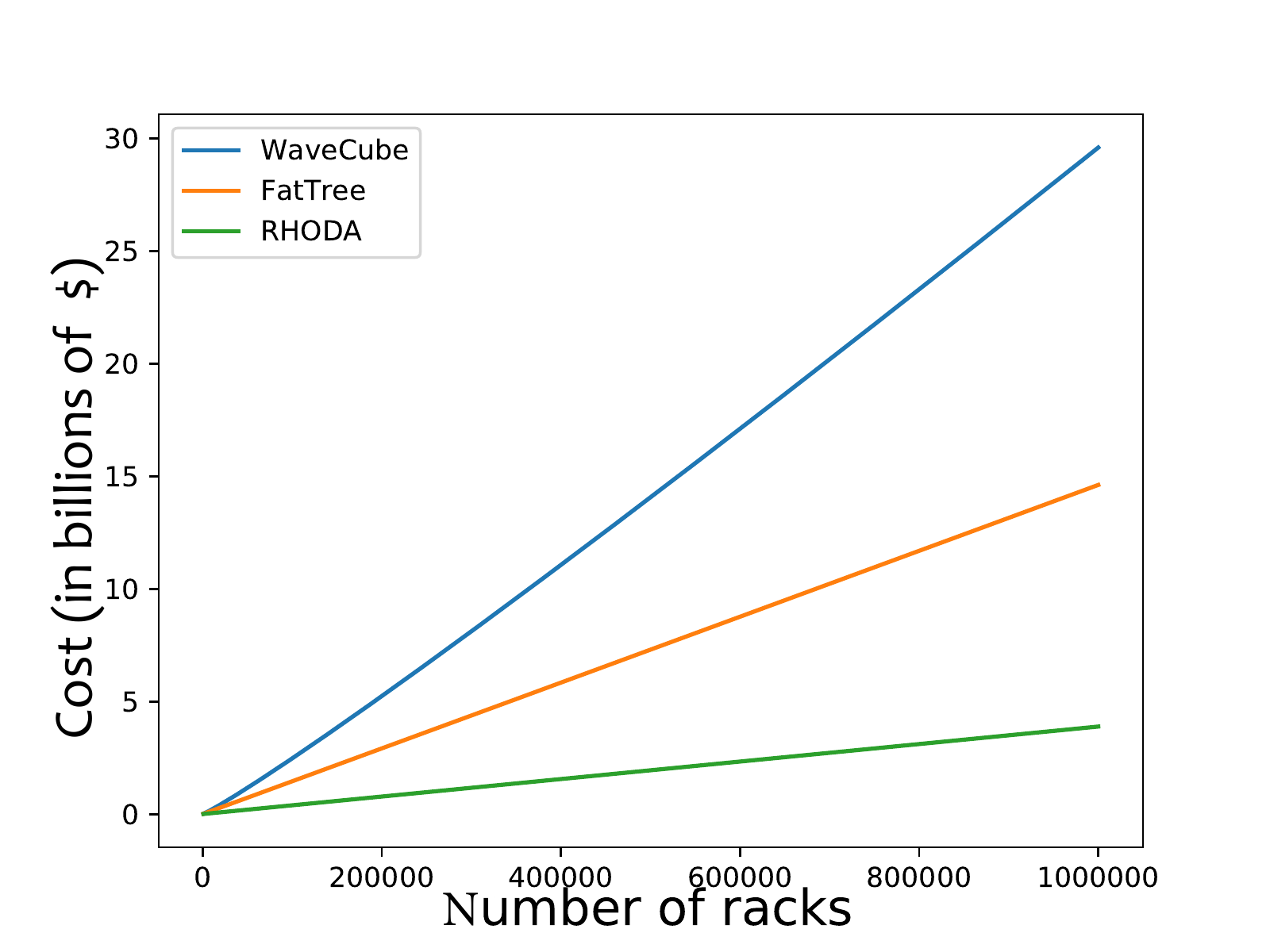}%
		\label{fig_c}%
	}%
	\vspace*{-.2cm}
	\caption{Power and dollar cost comparison.}
	\vspace{-10pt}
\end{figure}

For inter-cluster network, grooming switches are used to multiplex and pack data from different wavelengths. The average load of a grooming switch equals the amount of traffic from a cluster multiplied by the average number of hops in the inter-cluster network, i.e.,
\begin{equation*}
	T_{GS} = \frac{\alpha{\cdot}R}{M-1}{\cdot}(M-k){\cdot}k{\cdot}\bar{H}_{inter}.
\end{equation*}

So, the number of transmitters needed on all grooming switches is 
\begin{equation*}
N_{GS} = C{\cdot}\max(T_{GS}/100, 4),
\end{equation*}
where $4$ is the degree $d$ of the inter-cluster logical topology, and $100$ is the capacity of a wavelength in Gbps.

Each optical transceivers is assumed to operate at 100 Gbps, which is also the wavelength capacity. Each rack sends $R=640$ Gbps ($64\times 100$) and $\alpha=10\%$ is destined to the other $M-1$ racks. 
We calculate the power consumption and cost of these network topologies by summing up the consumed power and dollar cost of each component. A summary of these individual components is shown in Table \ref{power_cost}. Since our purpose is to compare the differences between these DCN interconnect schemes, the table does not include the data center servers and ToR switches.

Fig.~\ref{fig_p} presents the overall power consumption for various number of racks in the DCN. Even though each rack in WaveCube can communicate with $\log_2M$ racks, such a scheme also causes large power consumption. As shown in Table~\ref{power_cost}, WSS per port has a large power consumption, compared with other optical devices. Each rack in WaveCube needs a $1{\times}{\log_2M}$ WSS, which results in enormous power consumption. As shown in Table~\ref{power_cost}, electronic devices have larger consumption than optical devices. So, FatTree also has large power consumption. Fig.~\ref{fig_c} presents the overall cost for various number of racks in the DCN. In terms of power consumption, RHODA can save up to $86\%$ and $93\%$ compared with FatTree and WaveCube, respectively. In terms of cost, RHODA can save up to $78\%$ and $86\%$, respectively.

\vspace{-5pt}
\section{Scalability}\label{sec:scalability}
We briefly comment on the scalability of RHODA. The number of ports of AWGR and OSW can scale up to 512~\cite{cheung2014ultra} and 1024~\cite{ChenOSA}, respectively. If a rack accommodates 64 servers, then RHODA can scale up to 30 million $(=64{\cdot}512{\cdot}1024)$ servers (not considering the wavelength limit).

\begin{figure}[t!]
	\centering
	\includegraphics[height=1.4in,width=2.6in]{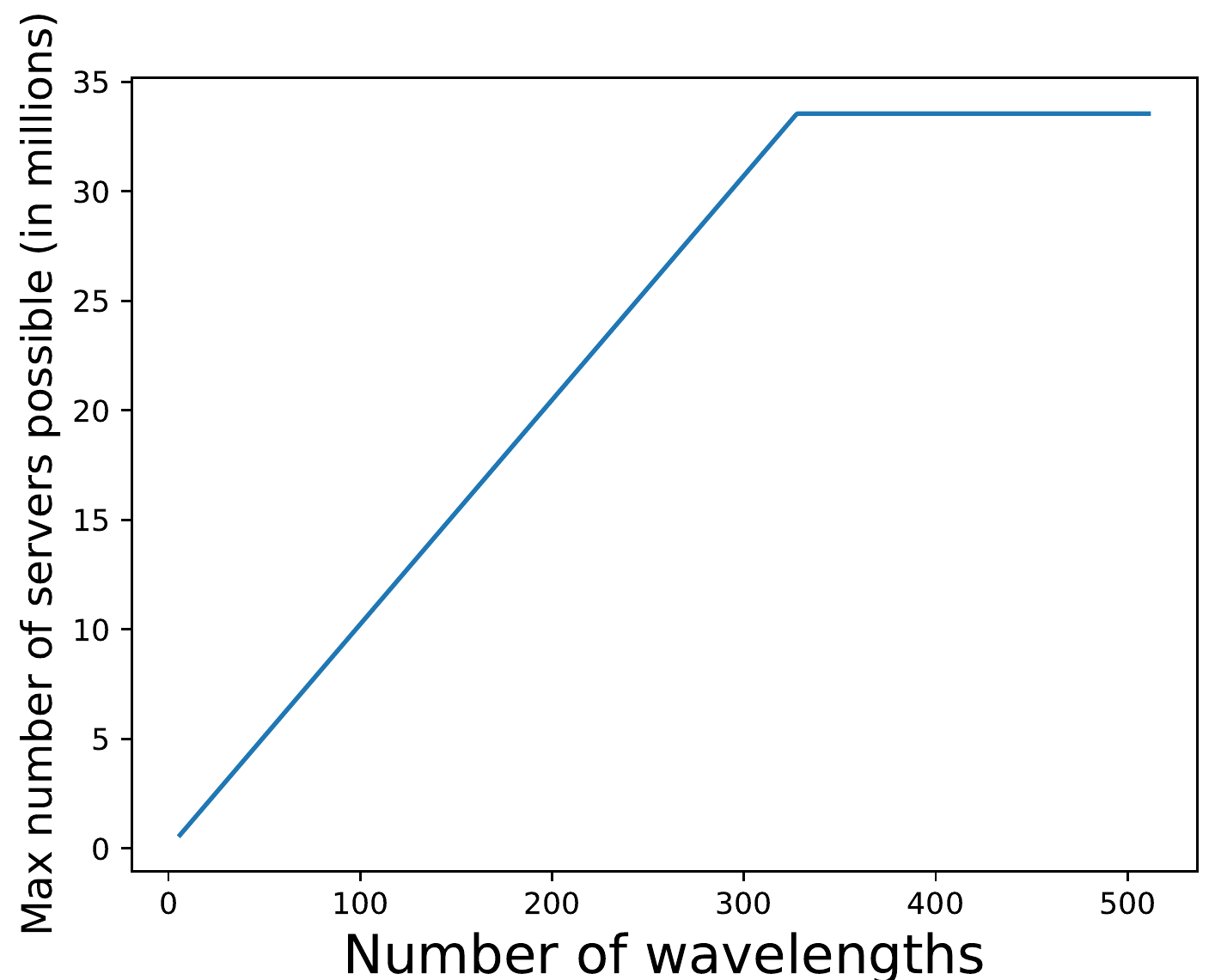}
	\caption{The scalability of RHODA.}
	\label{fig_scalability}
	\vspace{-15pt}
\end{figure}

Considering a limited number of wavelengths, if each server transmits at a maximum rate of 10 Gbps, and $90\%$ of traffic is for racks in the same cluster, then the amount of traffic that is sent out of a cluster is $64{\cdot}k$ Gbps. Suppose $W$ wavelengths are available and the capacity of each wavelength is 100 Gbps. Then, $W {\geq} \frac{640{\cdot}90\%}{100}$ for intra-cluster network. Also, for inter-cluster network $k {\leq} \min(\frac{100{\cdot}W}{64}, 512)$.  Figure~\ref{fig_scalability} shows the number of servers that can be accommodated as a function of the number of available wavelengths. For 128 wavelengths, the number of servers can reach up to 10+ million.


\vspace{-6pt}
\section{Conclusion}\label{sec:conclusion}
To accommodate massive numbers of servers and support variable traffic patterns, we introduce RHODA, a new two-level hierarchical and reconfigurable DCN architecture. Our simulation results demonstrate that, with cluster membership, inter-cluster topology, and bandwidth reconfigurability, RHODA can handle various traffic patterns, e.g., high fan-out, hotspots, etc. In terms of average hop distance, RHODA outperforms OSA, FatTre and WaveCube by up to $81\%$, $66\%$ and $60\%$, respectively. In future work, we will explore reconfigurability of RHODA at runtime, e.g., the ability of cluster membership, inter-cluster topology, and inter-cluster bandwidth reconfigurability at runtime. Also, we will explore approximation algorithms for routing and reconfiguration.




%

\bibliographystyle{IEEEtran}
\vspace{-5pt}
\bibliography{references}

\end{document}